\documentclass[prl,superscriptaddress,twocolumn,a4paper,showpacs]{revtex4}

\usepackage{amsfonts,amssymb,amsmath}
\usepackage[]{graphics,graphicx,epsfig}
\usepackage{amsthm}

\def\identity{\leavevmode\hbox{\small1\kern-3.8pt\normalsize1}}

\newcommand{\Tr}{\mathrm{Tr}}

\renewcommand{\epsilon}{\varepsilon}

\begin{document}

\title{Criteria for two distinguishable fermions to behave like a boson}

\author{Ravishankar \surname{Ramanathan}}
\affiliation{Centre for Quantum Technologies, National University of Singapore, 3 Science Drive 2, Singapore 117543}
\author{Pawel Kurzynski}
\affiliation{Centre for Quantum Technologies, National University of Singapore, 3 Science Drive 2, Singapore 117543}
\affiliation{Faculty of Physics, Adam Mickiewicz University, Umultowska 85, 61-614 Pozna\'{n}, Poland} 
\author{Tan Kok Chuan}
\affiliation{Centre for Quantum Technologies, National University of Singapore, 3 Science Drive 2, Singapore 117543}
\author{Marcelo F. Santos}
\affiliation{Departamento de F\'isica, Universidade Federal de Minas
Gerais, Belo Horizonte, Caixa Postal 702, 30123-970, MG, Brazil}
\author{Dagomir Kaszlikowski}
\email{phykd@nus.edu.sg}
\affiliation{Centre for Quantum Technologies, National University of Singapore, 3 Science Drive 2, Singapore 117543}
\affiliation{Department of Physics, National University of Singapore, 2 Science Drive 3, Singapore 117542}

%\author{Ravishankar Ramanathan$^{1}$, 
%Paul Kurzy\'nski$^{1,2}$,
%Tan Kok Chuan$^{1}$,
%Marcelo F. Santos$^{1,3}$
%and D Kaszlikowski$^{1,4}$}

%\address{
%  $^1$ Centre for Quantum Technologies, National University of Singapore, Singapore \\
%  $^2$ Faculty of Physics, Adam Mickiewicz University, Pozna\'{n}, Poland \\
%  $^3$ SUPA Department of Physics, University of Strathclyde, Glasgow G4 0NG, United Kingdom \\
%  $^4$ Department of Physics, National University of Singapore, Singapore}

%\ead{phykd@nus.edu.sg}

\begin{abstract}
We study the necessary conditions for bosons composed of two 
distinguishable fermions to exhibit bosonic-like behaviour. We base 
our analysis on tools of quantum information theory such as 
entanglement and the majorization criterion for probability 
distributions. In particular we scrutinize a recent interesting 
hypothesis by C. K. Law in the Ref. Phys. Rev. A {\bf 71}, 034306 
(2005) that suggests that the amount of entanglement between the 
constituent fermions is related to the bosonic properties of the 
composite boson. We show that a large amount of entanglement does not 
necessarily imply a good boson-like behaviour by constructing an 
explicit counterexample. Moreover, we identify more precisely the role 
entanglement may play in this situation.
\end{abstract}

\maketitle

\section{Introduction}
It is well-known that particles containing an even number of fermions behave like bosons under certain experimental conditions. This fact has been confirmed in numerous experiments where Bose-Einstein condensates of atoms have been created. This is generally believed to occur at low densities, where the overlap of the fermionic wave functions can be neglected. In \cite{Law,Wot}, it was hypothesised that the amount of entanglement between the constituent fermions plays a role in the boson-like behaviour. In this paper we scrutinize this idea further and show that a large amount of entanglement is not sufficient. We refine it and identify more precisely the role entanglement plays in boson-like behaviour of bipartite systems of two species of distinguishable fermions.  

We focus on the simplest scenario of a composite boson (coboson) made of two constituent distinguishable fermions. An example of such a system is an exciton made of an electron-hole pair in semiconductors. The bosonic behavior of excitons, especially in the context of Bose-Einstein condensation has been studied (see for example \cite{BEC}), and it is commonly accepted that excitons behave like bosons when the density is low such that the Pauli exclusion principle and the Coulomb interaction can be neglected.

In general, the state of a single composite boson made of two distinguishable fermions of type A and B can be written as
\begin{equation}\label{composite}
|\psi\rangle_{AB}=\sum_{n}\sqrt{\lambda_n}a^{\dagger}_n b^{\dagger}_n|0\rangle, \nonumber
\end{equation}
where $a^{\dagger}_n(b^{\dagger}_n)$ creates particle A (B) in mode $n$, and $\lambda_{n}$ is the probability of occupation of mode n. The sum in the above equation is over all modes from 0 to $\infty$, however for ease of calculation, we can introduce a cutoff $d$ (justified on grounds of finite energy), which later will be taken to infinity. Standard anti-commutation rules apply, i.e., $\{K_n,K^{\dagger}_m\}=\delta_{nm}$ ($K=a,b$) and creation/anihilation operators for fermions of type A and B anti-commute.

The number of non-zero coefficients $\lambda_{n}$ is the Schmidt number of this state and as long as it is larger than one the state is entangled. The amount of entanglement between fermions A and B can be quantified for instance by the von Neumann entropy of the reduced density matrix of one of the subsystems, which is simply the Shannon entropy of the probability distribution generated by $\lambda_n$, i.e., $-\sum_n\lambda_n\ln{\lambda_n}$.

A necessary condition for the composite system described by the state $|\psi\rangle_{AB}$ to exhibit boson-like behaviour is that the creation operator of the coboson
\begin{equation}\label{c}
c^{\dagger}=\sum_n \sqrt{\lambda_n} a^{\dagger}_n b^{\dagger}_n \nonumber
\end{equation}
mimics a proper bosonic creation operator as faithfully as possible. The corresponding commutation relation is given by
\begin{equation}
[c,c^{\dagger}]= \textbf{1} - \sum_n \lambda_n( a^{\dagger}_n a_n +  b^{\dagger}_n b_n). \nonumber
\end{equation}
The operator $\Delta=\sum_n \lambda_n( a^{\dagger}_n a_n +  b^{\dagger}_n b_n)$ can be interpreted as a deviation from the proper bosonic commutation relations and it should be as small as possible in some sense.

An interesting method of quantifying the smallness of $\Delta$ was investigated in \cite{CT}. Two operators were considered, the bosonic departure $1-[c,c^{\dagger}]$ and the boson number $c^{\dagger}c$, whose expectation value in the cobosonic number states $|N\rangle$ were postulated to be a measure of the {\it quality} of bosons. 

The state $|N\rangle$ is defined as follows
\begin{equation}\label{state1}
|N\rangle=\chi_N^{-1/2}\frac{c^{\dagger N}}{\sqrt{N!}}|0\rangle \nonumber.
\end{equation}
Note that the above state resembles a usual bosonic number state except for the normalization factor $\chi_N$, which is given by
\begin{equation}
\chi_{N} = N! \sum_{n_1< n_2 < \dots <n_N} (\lambda_{n_1} \lambda_{n_2} \dots \lambda_{n_N}) \nonumber
\end{equation}
The behaviour of the state $|N\rangle$ under the action of the corresponding annihilation operator $c$ is given by
\begin{equation}
c|N\rangle=\alpha_N \sqrt{N}|N-1\rangle+|\varepsilon_N\rangle. \nonumber
\end{equation}
The parameter $\alpha_N=\sqrt{\chi_N/\chi_{N-1}}$ is a normalization constant and
$|\varepsilon_N\rangle$ is a vector orthogonal to $|N-1\rangle$ of norm
\begin{equation}\label{eps}
\langle\varepsilon_N|\varepsilon_N\rangle=1-N\frac{\chi_{N}}{\chi_{N-1}}+(N-1)\frac{\chi_{N+1}}{\chi_N}. \nonumber
\end{equation}

The expectation values of the bosonic departure and boson number operators in $|N\rangle$ are
\begin{equation}
\left\langle 1-[c,c^{\dagger}] \right\rangle_{N} = 2\left(1-\frac{\chi_{N+1}}{\chi_{N}}\right) \nonumber
\end{equation}
and 
\begin{equation}
\left\langle c^{\dagger}c \right\rangle_{N} = N - \frac{N-1}{2}\left\langle 1-[c,c^{\dagger}] \right\rangle_{N} \nonumber
\end{equation}
As can be seen from the above equations, the ratio $F_{N+1} = \frac{\chi_{N+1}}{\chi_{N}}$ is a mathematical indicator of the quality of the boson. In particular, bosonic behavior is obtained when the ratio approaches one. Moreover, it can be seen that $\alpha_N\rightarrow 1$ and $\langle\varepsilon_N|\varepsilon_N\rangle \rightarrow 0$ as $\chi_{N + 1}/\chi_N \rightarrow 1$. 

In \cite{Law}, it was shown that for a particular (exponential) 
distribution of $\lambda_{k}$, the above ratio is in fact related to 
the entanglement between the constituent fermions and approaches one 
in the limit of infinite entanglement. Moreover, since in principle entanglement does not depend on the distance between the fermions, it was speculated that highly delocalized cobosons could be prepared and made to condense.

This result was extended in \cite{Wot}, where it was shown that the ratio $\chi_{N+1}/\chi_N$ can be bounded from below and from above by simple functions of the purity $P(\rho)=\Tr\{\rho_{A(B)}^2\}$ of the reduced density matrix $\rho_{A(B)}$ of particle A (B), which is an entanglement measure for pure states of the bipartition A-B. The result was
\begin{equation}\label{wt}
1-NP\leq \frac{\chi_{N+1}}{\chi_N}\leq 1 - P. 
\end{equation}
For highly entangled states, i.e., for which $P\rightarrow 0$, $\chi_{N+1}/\chi_{N}\rightarrow 1$. The above bounds were proposed to validate the hypothesis that the cobosonic behavior depends on the entanglement between A and B. 

Before proceeding with the main argument, let us show that one can obtain better approximations to the ratio $\frac{\chi_{N+1}}{\chi_{N}}$ by making use of the Newton-Girard identities that relate elementary symmetric polynomials $\chi_N$ to the complete symmetric polynomials $P_{j} = \sum_{m=0}^{\infty} \lambda_{m}^{j}$. This relation is given by
\begin{equation}
\chi_{N} = \sum_{j=1}^{\infty}N^{j-1}P_j\chi_{N-j}. \nonumber
\end{equation}
Note that $P_{2} = P$ is the purity defined earlier. 
Dividing the above equation throughout by $\chi_{N}$ and noting that for large N, $\frac{\chi_{N}}{\chi_{N-j}} \approx \left(\frac{\chi_{N}}{\chi_{N-1}}\right)^j$, we obtain 
\begin{equation}
1 \approx \frac{1}{F_{N}} - \frac{N P_{2}}{F_{N}^2} + \frac{N^2 P_{3}}{F_{N}^3} - \frac{N^3 P_{4}}{F_{N}^4} + \dots \nonumber.
\end{equation}
 
A necessary condition for good bosonic behaviour is that for all k, $\lambda_{k} \ll 1/N$, which implies that the terms $N^{j-1}P_j$ in the expression above decrease in magnitude sharply \cite{CBD}. For instance, for a typical $\lambda_{k}$  of the order of $O(1/N^2)$, the expression $N^{j-1} P_j$ is of the order of $O(1/N^{j-1})$. Hence, performing a series expansion around these small terms leads to
\begin{equation}
\frac{\chi_{N}}{\chi_{N-1}} \approx 1 - N P_{2} + N^{2}(P_{3} - P_{2}^{2}) + O(N^3 P_{2}^{3}). \nonumber
\end{equation} 
Better approximations to this important ratio for large N can be made by considering more terms in the series expansion above. 

However, we shall see that high entanglement between pairs of fermions is not sufficient to make them behave like bosons. In general, the flagship bosonic phenomena like bunching and Bose-Einstein condensation do not depend on the interaction between bosons. Therefore we consider a model in which there is no interaction between fermions A and B because these processes should be possible even in the absence of such interaction. In quantum information theory such processes are referred to as LOCC (local operations and classical communication) ones. In our case, this would imply local operations among fermions of type A and B only (see Figure \ref{f1}). A notable characteristic of these is that entanglement cannot be increased, which would require interaction between subsystems. 

The necessary and sufficient criterion for determinisitc LOCC transformation between bipartite quantum states was given in \cite{Nielsen} where the mathematical concept of majorization was introduced into quantum information theory. Since we want the cobosons to behave like real bosons, the deterministic LOCC process of formation of $N$-particle cobosonic states $|N\rangle$ must be possible. As a result, the corresponding final and initial cobosonic states have to obey the majorization criterion. We show that $P\rightarrow 0$ does not imply majorization, by providing a counterexample. Finally, we find a sufficient condition for bosonic behavior incorporating majorization and high entanglement.

\section{Condensation and LOCC conditions}

In this section we refer to the process of generating the state $|N\rangle$ from $N$ single-particle cobosonic states $|\psi\rangle_{AB}^{\otimes N}$ as a process of creating a condensate. It should be noted that this is a simplified notion of the actual picture, since we do not consider parameters such as temperature. However, this simplification captures the essential feature of the condensation process with regard to the macroscopic occupation of a single state. 

\begin{figure}
\begin{center}
\includegraphics[scale=0.35]{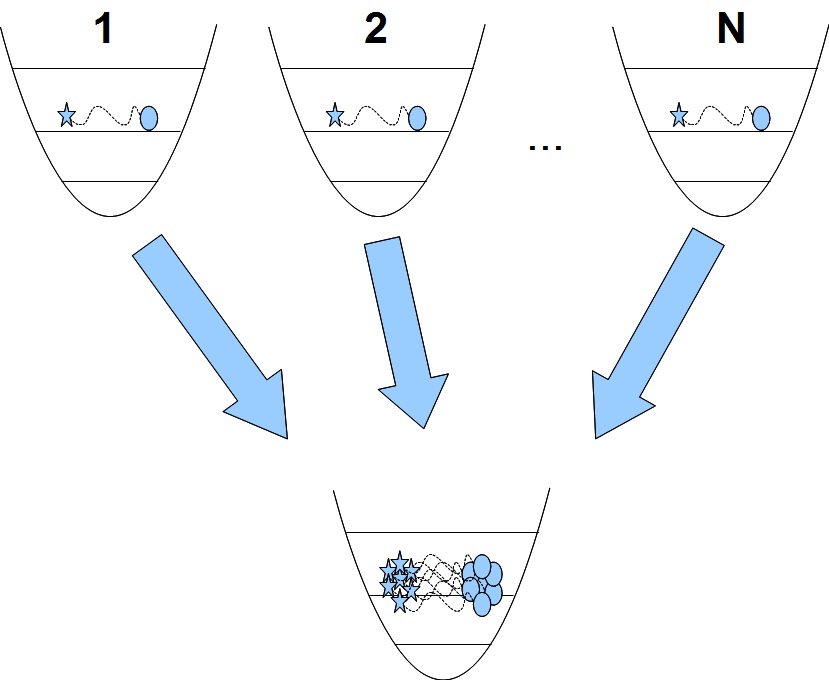}
\end{center}
\caption{The schematic picture presenting the idea of condensation of cobosons. Each coboson initially occupies a different potential well. The well energy levels enumerated by $n$ give rise to the internal structure of cobosons $c^{\dagger}=\sum_n\sqrt{\lambda_n}a_n^{\dagger} b_n^{\dagger}$. We are looking for the possibility of existence of an LOCC operation bringing all cobosons into one well.
}
 \label{f1}
\end{figure}

Let us start with $N$ identical cobosons in $N$ different potential wells as illustrated in Figure \ref{f1}.
The initial state is thus given by
\begin{equation}\label{init}
c_1^{\dagger}c_2^{\dagger}\dots c_N^{\dagger}|0\rangle, \nonumber
\end{equation}
where $c_j^{\dagger}=\sum_n \sqrt{\lambda_n} a_n^{(j)\dagger} b_{n}^{(j)\dagger}$ creates one coboson in the j'th well. Our goal is to condense all cobosons in a single well, i.e., to obtain the final state $(c^{\dagger})^{N}|0\rangle$ (up to normalization). Since we do not allow parts A and B to interact (LOCC), the process of condensation cannot increase the entanglement between A and B.  The allowed set of deterministic LOCC transformations on bipartite entangled states is given by the majorization criterion \cite{Nielsen}, which states that a state $|\psi\rangle_{AB}$ can be transformed by LOCC into another state $|\phi\rangle_{AB}$ iff the vector of eigenvalues of the density matrix of one of the two subsystems of state $|\psi\rangle_{AB}$, denoted by $\vec{\lambda}_{\psi}$, is majorized by the corresponding vector $\vec{\lambda}_{\phi}$ of state $|\phi\rangle_{AB}$. We write $\vec{\lambda}_{\psi}\prec\vec{\lambda}_{\phi}$, which means that for all $k=0,1,2,\dots,d-1,$
$\sum_{j=0}^{k}\lambda_j^{\downarrow}(\psi) \leq \sum_{j=0}^{k}\lambda_j^{\downarrow}(\phi)$, with equality $\sum_{j=0}^{d-1}\lambda_j^{\downarrow}(\psi) = \sum_{j=0}^{d-1}\lambda_j^{\downarrow}(\phi) = 1$, where $\downarrow$ symbol indicates that the eigenvalues are enumerated in decreasing order, i.e., $\lambda_0\geq \lambda_1 \geq \dots \geq \lambda_{d-1}$. In order to verify that majorization applies to fermionic systems, let us emphasize that all the necessary transformations can be performed while the cobosons occupy different wells. Hence, there exists a parameter which allows us to distinguish between different cobosons, and then to perform one-to-one mapping from $N$ wells into one. 

The majorization criterion has to be applied to the vectors of eigenvalues of the initial and final reduced density matrices of subsystem A or B ($\rho_{i}$ and $\rho_{f}$), which in our case are given by the diagonal matrices
\begin{eqnarray}
\rho_i &=& \sum_{n_1,n_2,\dots,n_N}\lambda_{n_1}\lambda_{n_2}\dots\lambda_{n_N} \times \nonumber \\
&\times& a_{n_1}^{(1)\dagger}a_{n_2}^{(2)\dagger}\dots a_{n_N}^{(N)\dagger}|0\rangle\langle 0|a_{n_N}^{(N)}\dots a_{n_2}^{(2)} a_{n_1}^{(1)}, \nonumber
\label{rhoin}
\end{eqnarray}
and
\begin{eqnarray}
\rho_f &=& \frac{1}{\tilde{\chi}_N}\sum_{n_1 < n_2 <\dots <n_N}\lambda_{n_1}\lambda_{n_2}\dots\lambda_{n_N} \times \nonumber \\
&\times& a_{n_1}^{\dagger}a_{n_2}^{\dagger}\dots a_{n_N}^{\dagger}|0\rangle\langle 0|a_{n_N}\dots a_{n_2} a_{n_1}, \nonumber 
\label{rhof}
\end{eqnarray}
where $\tilde{\chi}_N=\sum_{n_1 <\dots < n_N} \lambda_{n_1} \dots \lambda_{n_N}$ is a normalization factor which is related to $\chi_N$ via  $\chi_N=N! \tilde{\chi}_N$. In the second equation we skipped the superscript $(j)$, since all operators correspond to cobosons in one well. In this sense, condensation is a process in which the cobosons lose their identity and become indistinguishable once all of them are in the same well. Since the $\lambda$'s are enumerated decreasingly, the ordered vectors of eigenvalues for the initial ($\vec{\lambda}_{i}$) and final ($\vec{\lambda}_{f}$) states are $\vec{\lambda}_i=\{\lambda_0^N, \lambda_0^{N-1}\lambda_1,\dots\}$, and $\vec{\lambda}_f=\frac{1}{\tilde{\chi}_{N}}\{\lambda_0\lambda_1\dots\lambda_{N-1},\dots\}$.

Note, that in this case the vector $\vec{\lambda}_i$ is $d^{N}$-dimensional and $\vec{\lambda}_f$ is $\begin{pmatrix} d \\ N \end{pmatrix}$-dimensional. The dimension of the second vector being smaller than that of the first, we augment $\vec{\lambda}_f$ by $d^N - \begin{pmatrix} d \\ N \end{pmatrix}$ zeros. In the following sections, we shall look for cobosonic states described by distributions $\{\lambda\}$ for which $\vec{\lambda}_i \prec \vec{\lambda}_f$.

\section{Large entanglement does not imply majorization}

We show that small purity $P$ does not imply majorization. We find cobosonic states for which the purity is small but vectors $\vec{\lambda}_i$ and $\vec{\lambda}_f$ corresponding to $|\psi\rangle_{AB}^{\otimes N}$ and $|N\rangle$, respectively, do not obey the majorization criterion. For simplicity, we consider the violation of the majorization relation between the first elements of these vectors
\begin{equation}\label{majbreak}
\lambda_0^N > \frac{\lambda_0\dots \lambda_{N-1}}{\tilde{\chi}_N}.
\end{equation}
Noting that $\lambda_{j}'s$ are arranged in descending order, we can write $\lambda_j := \lambda_0 \gamma_j$, with $0\leq \gamma_j \leq 1$, $\gamma_{j+1}\leq \gamma_j$ and $\gamma_0 = 1$. Therefore, Eq. (\ref{majbreak}) can be rewritten as $\tilde{\chi}_N > \gamma_1 \gamma_2 \dots \gamma_{N-1}$. Using (\ref{wt}), one finds that 
\begin{equation}
\tilde{\chi}_N\geq\tilde{\chi}_{N-1}\left(\frac{1+P}{N}-P\right). \nonumber
\end{equation}
Applying the above relation $N$ times we obtain
\begin{equation}
\tilde{\chi}_N\geq \frac{\tilde{\chi}_1}{N!}\prod_{j=2}^N(1+P-jP) \geq \frac{1}{N!}\left(1-(N-1)P\right)^{N-1}, \nonumber
\end{equation}
therefore majorization fails if
\begin{equation}\label{cond}
\gamma_1\gamma_2\dots \gamma_{N-1}< \frac{1}{N!}\left(1-(N-1)P\right)^{N-1}.
\end{equation}

Next, let us consider cobosonic states with
\begin{equation}\label{riemann}
\lambda_{j}=\frac{1}{(j+1)^s \zeta(s)}, 
\end{equation}
where 
\begin{equation}
\zeta(s)=\sum_{j=0}^{\infty} \frac{1}{(j+1)^{s}} \nonumber
\end{equation}
is the Riemann Zeta function. For $s>1$ the sum of the $\lambda_j$-series is one, thus it is a valid probability distribution. The purity for these states is given by
\begin{equation}\label{purity}
P_s=\frac{\zeta(2s)}{\zeta(s)^2}. 
\end{equation}
It is small for $s=1+\varepsilon$ ($0 < \varepsilon \ll 1$), since one can approximate (\ref{purity}) by
\begin{equation}\label{purityapprox}
P_{1+\varepsilon}\approx\zeta(2)\varepsilon^2=\frac{\pi^2}{6}\varepsilon^2. \nonumber
\end{equation}
Moreover, in this case $\gamma_j=\frac{1}{(j+1)^{1+\varepsilon}}$. Plugging this into (\ref{cond}) gives
%\begin{equation}
%\frac{1}{N!^{1+\varepsilon}}< \frac{1}{N!}[1-(N-1)P_{1+\varepsilon}]^{N-1}, \nonumber
%\end{equation}
%or equivalently
\begin{equation}\label{violation}
N!^{\varepsilon}[1-(N-1)P_{1+\varepsilon}]^{N-1}>1. \nonumber
\end{equation}
For $N \varepsilon \leq 1$, we can expand the above into power series and keep only the terms up to the first order in $\varepsilon$
\begin{equation}\label{aproxviolation}
1+ \varepsilon \log(N!)>1. \nonumber
\end{equation}
It is clear that the majorization condition is violated. Let us also point out that despite the fact that condensation is not possible by LOCC, the ratio $F_{N} = \frac{\chi_{N}}{\chi_{N-1}}$ can be  close to one. Taking into account the lower bound in (\ref{wt}), we have
\begin{equation}\label{frac}
F_N\geq 1 - \frac{\pi^2}{6}N\varepsilon^2. \nonumber
\end{equation}
The right hand side is close to one for $N\varepsilon^2 \ll 1$. This shows that majorization is a qualitatively different condition for bosonic behavior than previously studied. 

\section{Sufficient condition for condensation}

In the references \cite{ExBEC1, ExBEC2}, it was argued that a necessary condition for bosonic behavior of cobosons (in particular, of excitons) is 
\begin{equation}\label{dens}
N\lambda_0\ll 1.
\end{equation}
This condition is related to the fact that the density of cobosons has to be low in order to prevent overlap between wavefunctions of underlying fermions. 

We now show that this condition is more general in that it implies $N P \ll 1$, but not vice versa. This can be seen from the fact that the purity is bounded from above by the largest coefficient $\lambda_0$. Since $\sum_j \lambda_j = 1$, we have $\sum_j \gamma_j = \frac{1}{\lambda_0}$ and the following holds
\begin{equation}\label{b1}
P=\sum_j\lambda_j^2=\lambda_0^2\sum_j\gamma_j^2 \leq \lambda_0^2\sum_j \gamma_j =\lambda_0. \nonumber 
\end{equation}
Therefore $P \ll \frac{1}{N}$ is guaranteed if $\lambda_0 \ll \frac{1}{N}$. 

However, the converse is not true ($NP \ll 1$ does not imply $N \lambda_{0} \ll 1$). To see this, let us consider the states (\ref{riemann}) for $s=1+\varepsilon$. For $\varepsilon=\frac{1}{N}$ the purity is $O(1/N^2)$, but the condition (\ref{dens}) is not satisfied, because $\lambda_0=\frac{1}{\zeta(1+\varepsilon)}\approx \varepsilon$ and $N\lambda_0\approx 1$. 

In general, it is hard to give a necessary and sufficient condition for the majorization of the initial Schimidt vector of N cobosons $\vec{\lambda}_{i}$ by the final Schmidt vector of the N coboson condensate $\vec{\lambda}_{f}$. However, we would like to give a sufficient condition for majorization that has a simple physical interpretation. To do so, let us begin by considering the situation when $d$ and $N$ are finite, and then take the limit $d \rightarrow \infty$ and $N \rightarrow \infty$ while keeping the ratio $\mu = \frac{N}{d}$ constant. $\mu$ is a quantity proportional to the density of cobosons $\rho = N/V$ since $\mu$ =  $\frac{\rho}{\omega}$, where $\omega$ is the density of states. 

Let the state of our cobosons be described by the vector $\vec{\lambda} = \left\{\lambda_{0}, \dots, \lambda_{d}\right\}$. Firstly, let us note that if it were possible, by LOCC, to transform $\vec{\lambda}_{i}$ to a vector $$\vec{u}_{f}=\left\{\begin{pmatrix} d \\ N\end{pmatrix}^{-1}, \dots, \begin{pmatrix} d \\ N\end{pmatrix}^{-1}\right\},$$ it would be possible to transform by LOCC $\vec{\lambda}_{i}$ to $\vec{\lambda}_{f}$ since the vector $\vec{u}_{f}$ corresponds to the uniform distribution in the final Hilbert space and hence is guaranteed to be majorized by $\vec{\lambda}_{f}$. Note that the distribution $\vec{u}_{f}$ corresponds to the state with the highest $F_{N}$ and is therefore, the most ideal cobosonic condensate. The majorization problem $\vec{\lambda}_{i} \prec \vec{u}_{f}$ yields a simple condition $\lambda_{0}^N \leq \begin{pmatrix} d \\ N\end{pmatrix}^{-1}$ which in the infinite limit translates simply to 
\begin{equation}
\lambda_{0} \leq \mu \nonumber.
\end{equation}
The above condition, in the spirit of \cite{CT, ExBEC1}, provides a sufficient condition for the bosonic behavior of these particles, to wit that the probability of occupation of the ground state be less than the occupational density. It also supports the idea of a critical density for condensation, that is, the equation above may be satisfied for all densities above a critical value $\mu_{cr}$, ensuring that such distributions lead to a condensation without interaction between the fermions as required for proper bosonic behavior. 

Now, we show that there exist certain classes of states that always obey the majorization condition. Let us consider states 
\begin{equation}\label{laws}
\lambda_{j}=(1-z)z^j,~~0<z<1,
\end{equation}
which were introduced in \cite{Law}. For these,
\begin{equation}
\tilde{\chi}_N=\frac{z^{N(N-1)/2}(1-z)^N}{\prod_{j=1}^N (1-z^j)}, \nonumber
\end{equation}
and the purity is given by
\begin{equation}
P_z=\frac{1-z}{1+z}. \nonumber
\end{equation}
For $z \rightarrow 1$ the purity is small. Below we show that these states obey the majorization criterion for all z and N. 

Firstly, observe that
\begin{equation}
\vec{\lambda}_i=(1-z)^N\{1,z,z,\dots,z,z^2,z^2,\dots\}, \nonumber
\end{equation}
where the degeneracy of $z^l$ is $g_{l}^{(i)} = \begin{pmatrix} l+N-1 \\ l\end{pmatrix}$. On the other hand, 
\begin{equation}
\vec{\lambda}_f=\frac{(1-z)^N z^{N(N-1)/2}}{\tilde{\chi}_N}\{1,z,z^2,z^2,\dots\}, \nonumber
\end{equation}
for which the degeneracy of  $z^l$ ($g_{l}^{(f)}$) is smaller than $g_{l}^{(i)}$. The majorization states that, for all $k$, the sum of the first $k$ terms of $\vec{\lambda}_i$ has to be less than, or equal to the sum of the first $k$ terms of $\vec{\lambda}_f$. Canceling common factors, the overall multiplicative factor in $\vec{\lambda}_f$ is 
\begin{equation}
\frac{z^{N(N-1)/2}}{\tilde{\chi}_N}= \prod_{k=1}^{N-1} \sum_{j=0}^{k} z^j, \nonumber
\end{equation}
and we can rewrite
\begin{equation}
\vec{\lambda}_i'=\{1,z,z,\dots,z,z^2,z^2,\dots\},\nonumber
\end{equation}
and
\begin{equation}
\vec{\lambda}_f'= \left(\prod_{k=1}^{N-1} \sum_{j=0}^{k} z^j\right) \{1,z,z^2,z^2,\dots\}.\nonumber
\end{equation}
Next, notice that the sum over all terms of both vectors is the same and equal to $1/(1-z)^N$. Moreover, this sum can be written as a unique polynomial 
\begin{equation}
\frac{1}{(1-z)^N}=\sum_{j=0}^{\infty} g_{j}^{(i)} z^{j}.\nonumber
\end{equation}
 The sum of first $k$ terms of $\vec{\lambda}_i'$ is given by
\begin{equation}
\sum_{j=1}^{k}[\vec{\lambda}_i']_j= \sum_{j=0}^{l-1} g_{j}^{(i)} z^{j} + (k - \sum_{j=0}^{l-1} g_{j}^{(i)}) z^{l}.\nonumber
\end{equation} 
Due to the multiplicative factor of $\vec{\lambda}_f'$ and the fact that the degeneracy of $\vec{\lambda}_f'$ is less than that of $\vec{\lambda}_i'$, we have
\begin{equation}
\sum_{j=1}^{k}[\vec{\lambda}_i']_j < \sum_{j=1}^{k}[\vec{\lambda}_f']_j \nonumber
\end{equation}
which ends the proof.
We have shown that the states in (\ref{laws}) obey the majorization criterion for any $z$ and $N$. As a consequence, the corresponding cobosons should indeed behave like ideal bosons for large $z$ (when $P$ is small).

We now note that there may exist certain classes of states which do not describe proper cobosons, in the sense that they do not obey the majorization relation. However, there exists the interesting possibility that some of these states are transformable by LOCC into states which obey majorization (such as in (\ref{laws}) for large z) and therefore can be converted into proper cobosonic states. This means that they can be converted into condensates of different coboson type, i.e., $$c \rightarrow c' \rightarrow c'^{N}.$$

A sufficient criterion for majorization \cite{Maj} can be stated as:
If there exists an $i$ $(1\leq i \leq n)$ such that for all $k \leq i$, $\sum_{j=1}^{k} x_{j} \leq \sum_{j=1}^{k} y_{j}$, and for $k > i$, $\sum_{j=i+1}^{n} x_{j} \geq \sum_{j=i+1}^{n} y_{j}$, then $\vec{x} \prec \vec{y}$. This implies that any distribution of $\lambda's$ with $\lambda_{0} \leq (1-z)$ (with $z \rightarrow 1$) and with only one intersection with the exponential distribution in (\ref{laws}) will be majorized by the latter, and can be considered as a good cobosonic distribution. 

\section{Conclusions}
In this article, we have studied the conjecture made in \cite{Law} that the amount of entanglement between the constituent fermions of a composite boson is related to its bosonic properties. We note that while necessary, entanglement may not be sufficient to ensure ideal bosonic behavior of these states. Our studies were made in relation to the majorization condition in the context of creating a cobosonic number state $|N\rangle$ from $N$ single particle states $|\psi\rangle_{AB}^{\otimes N}$. 

We provide a counter-example in terms of the Riemann distribution which shows that the property of small purity does not necessarily lead to majorization. We also note that a more fundamental condition is that given in terms of small density, to wit, $N \lambda_{k} \ll 1$ which leads to small purity, but not vice versa. Also, we introduced a sufficient condition for majorization which lends support to the idea of a critical density for condensation. We note that certain states such as those in \cite{Law} always obey majorization. This leads to the interesting possibility that there may exist certain states, which while not obeying majorization themselves, might be transformable by LOCC to states that do. 

%Note that while we have been studying majorization as a condition for nonlocal cobosons, it is also essential for local cobosons in the absence of interaction. That is, 

Lack of majorization implies the necessity of interaction for the process of condensation to occur. For instance, an interesting future problem would be to study the excitonic states in \cite{CT} in the context of majorization, to check if excitonic condensates can be prepared in the absence of interaction between electrons and holes. 

{\em Acknowledgements}.  This research is supported by the National Research Foundation and Ministry of Education in Singapore.
We acknowledge useful discussions with Rosario Fazio, Tomasz Paterek and John Goold.

\end{document}